\documentclass[12pt]{article}
\usepackage{float} 
\usepackage{amsmath}
\usepackage{slashed}
\usepackage{amsfonts}   
\usepackage{amssymb}

\begin{document}
\date{}
\title{{\bf{\Large Lax pairs for string Newton Cartan geometry}}}
\author{
 {\bf {\normalsize Dibakar Roychowdhury}$
$\thanks{E-mail:  dibakarphys@gmail.com, dibakarfph@iitr.ac.in}}\\
 {\normalsize  Department of Physics, Indian Institute of Technology Roorkee,}\\
  {\normalsize Roorkee 247667, Uttarakhand, India}
\\[0.3cm]
}

\maketitle
\begin{abstract}
In this paper, based on a systematic formulation of Lax pairs, we show \textit{classical} integrability for nonrelativistic strings propagating over \textit{stringy} Newton-Cartan (NC) geometry. We further construct the corresponding \textit{monodromy} matrix  which leads to an infinite tower of \textit{non-local} conserved charges over stringy NC background.
\end{abstract}
\section{Overview and Motivation}
During last one decade, nonrelativistic strings coupled to background fields has started to unveil a new fascinating sector of string theory whose dynamics is governed by a 2D relativistic non-linear sigma model \cite{Gomis:2000bd}-\cite{Harmark:2017rpg} that enjoys a deformed \textit{global} string Galilei invariance on the world-sheet. The corresponding target space geometry\footnote{Also known as the \textit{string} Newton-Cartan (NC) background.} is obtained by gauging the \textit{extended} string Galilei algebra which eventually turns out to be an extension from point particle Newton-Cartan (NC) background to strings. As a natural consequence of this gauging procedure, the corresponding target space geometry is naturally attributed to both longitudinal ($ \tau_{M}^{i} $) as well as transverse ($ e_{M}^{q} $) vielbein fields which together with the $ U(1) $ connection ($ m_{M}^{i} $) define what is known as the stringy NC background \cite{Andringa:2012uz},\cite{Bergshoeff:2018yvt}. From the perspective of a 2D world-sheet theory, these gauge fields on the target space have a natural interpretation of being the coupling constants for the non-linear sigma model in the so called NC limit \cite{Andringa:2012uz}. 

Given the above structure, the purpose of the present analysis is to look ahead one step further and address the following fundamental question namely whether nonrelativistic strings propagating over string NC geometry are \textit{integrable}\footnote{This has always been an important question to pursue as this guarantees the exact solvability of string spectrum in its non relativistic formulation. This further opens up new exciting directions from the perspective of gauge/string duality. For example, if we claim that these nonrelativistic strings are dual to certain operators in gauge theory then the present analysis strongly points towards a new class of integrable QFTs whose spectrum should differ from that of $ \mathcal{N}=4 $ SYM \cite{Minahan:2002ve}.}. In other words, whether the non linear sigma model (characterizing the string propagation over stringy NC geometry) allows infinite symmetry generators associated to a one parameter family of flat connections those are constructed by means of certain matrix valued fields corresponding to generators of extended string Galilei algebra \cite{Andringa:2012uz}. Technically speaking, the present analysis is the nonrelativistic counterpart of the earlier observations made in the context of $ AdS_5 \times S^5 $ superstrings\footnote{See \cite{Bena:2003wd}-\cite{Beisert:2010jr} and references therein.}.

The present analysis is based on the construction of \textit{Lax} pairs using Maurer-Cartan form \cite{Grosvenor:2017dfs} corresponding to generators of extended string Galilei algebra \cite{Andringa:2012uz}. Based on the notion of spacetime foliation \cite{Bergshoeff:2018yvt}, we split this one form into longitudinal,
\begin{eqnarray}
\mathfrak{j}_{\alpha}^{(L)}=\mathfrak{j}_{\alpha}^{(\tau)i}\mathcal{H}_{i}+\mathfrak{j}_{\alpha}^{(m)i}\mathcal{Z}_{i}~;~i=0,1\label{ref1} 
\end{eqnarray}
as well as transverse 
\begin{eqnarray}
\mathfrak{j}_{\alpha}^{(T)}=\mathfrak{j}^{(e)q}_{\alpha}\mathcal{P}_{q}~;~q=2,..,d
\label{ref2}
\end{eqnarray}
components where $ \mathcal{H}_{i} $ and $ \mathcal{P}_{q} $ are respectively the generators corresponding to longitudinal as well as transverse translations. On the other hand, $ \mathcal{Z}_{i} $ is the non central extension of the string Galilean algebra\cite{Andringa:2012uz}.

As a first step towards proving integrability, we show that the non-linear sigma model under consideration could be expressed as a \textit{bilinear} in (\ref{ref1}) and (\ref{ref2}) whose equations of motion could be reproduced by means of a one parameter family of flat connections. These are precisely the Lax connections for the sigma model under consideration that could be expanded in a basis of string Galilei generators $ \lbrace\mathcal{H}_{i}, \mathcal{Z}_{i}, \mathcal{P}_{q}\rbrace $. Finally, using Lax connection(s), we construct the so called \textit{monodromy} matrix that leads to an infinite tower of non-local charges thereby proving integrability of the non-linear sigma model.

The rest of the paper is organized as follows. In Section 2, we provide a detailed analysis on the construction of Lax pairs for the nonrelativistic strings propagating over stringy NC background. In Section 3, we construct the corresponding tower of non-local charges. Finally, we conclude in Section 4. 
\section{Lax pairs}
We start our analysis with a formal introduction to the Nambu-Goto (NG) strings in NC background. The corresponding NG action associated with the \textit{bosonic} sector could be formally expressed as\footnote{For the purpose of our present analysis it is sufficient to set the dilaton equal to zero.}\cite{Bergshoeff:2018yvt},
\begin{eqnarray}
\mathcal{S}_{NG}=-\tau_{S} \int d^{2}\xi ~\sqrt{-| \mathcal{G}|}+ \tau_{S} \int\mathcal{B}^{(2)} ~;~\tau_{S} =\frac{1}{2 \pi \alpha'}\label{e1}
\end{eqnarray}
where, we read off the pull-back of the target space metric as,
\begin{eqnarray}
\mathcal{G}_{\alpha \beta}=\mathcal{G}_{MN}\partial_{\alpha}X^{M}\partial_{\beta}X^{N}
\end{eqnarray}
together with, $ X^{M}(M =0,..,d) $ as embedding coordinates and $ \alpha =\sigma^{0} , \sigma^{1}$ being the coordinates on the world-sheet. Here, $ B^{(2)} $ is the background NS-NS two form field which the string is coupled to.

Following \cite{Andringa:2012uz}, we express the metric field as,
\begin{eqnarray}
\mathcal{G}_{MN}=\mathfrak{E}^{a}_{M}\mathfrak{E}^{b}_{N}\vartheta_{ab}=\mathfrak{E}^{i}_{M}\mathfrak{E}^{j}_{N}\eta_{ij}+\mathfrak{E}^{p}_{M}\mathfrak{E}^{q}_{N}\delta_{pq}\label{e3}
\end{eqnarray}
where, $ i,j=0,1 $ are the \textit{longitudinal} directions together with $ p,q=2,..,d $ as being that of the \textit{transverse} directions associated with the tangent space $ \mathcal{T}_{p} $ at a point $ p $ in the string NC manifold ($ \mathcal{M} $). Here, $\mathfrak{E}^{a}_{M}  $ is a $ d+1 $ dimensional \textit{relativistic} vielbein that could be splited into following two parts \cite{Andringa:2012uz}, 
\begin{eqnarray}
\mathfrak{E}^{i}_{M}=\omega \tau^{i}_{M}+\frac{1}{ 2\omega}m_{M}^{i}~;~~\mathfrak{E}^{p}_{M}=e_{M}^{p}
\end{eqnarray}
where, $ \tau^{i}_{M} $ and $ e_{M}^{p} $ are respectively the \textit{longitudinal} as well as \textit{transverse} vielbein fields associated with the corresponding generators of translation \cite{Andringa:2012uz},\cite{Bergshoeff:2018yvt}. Here, $ \omega $  is the so called NC parameter that needs to be set to infinity for a consistent non relativistic limit \cite{Andringa:2012uz}. Finally, we introduce $ m_{M}^{i} $ as $ U(1) $ connection corresponding to the generator associated with the \textit{non central} extension of the string Galilei algebra \cite{Bergshoeff:2018yvt}.

Using (\ref{e3}), it is now straightforward to show,
\begin{eqnarray}
\mathcal{G}_{\alpha \beta}=\omega^{2}\tau^{i}_{\alpha}\tau^{j}_{\beta}\eta_{ij}+\mathfrak{H}_{\alpha \beta}+\frac{1}{4\omega^{2}}m_{\alpha}^{i}m_{\beta}^{j}\eta_{ij}
\end{eqnarray}
where, we introduce the so called transverse metric \cite{Bergshoeff:2018yvt},
\begin{eqnarray}
\mathfrak{H}_{MN}=e_{M}^{p}e_{N}^{q}\delta_{pq}+\frac{1}{2}(\tau_{M}^{i}m_{N}^{j}+m_{M}^{i}\tau_{N}^{j})\eta_{ij}~;~\mathfrak{H}_{\alpha \beta}:=\mathfrak{H}_{MN}\partial_{\alpha}X^{M}\partial_{\beta}X^{N}
\end{eqnarray}
together with,
\begin{eqnarray}
\tau^{i}_{\alpha}= \tau^{i}_{M}\partial_{\alpha}X^{M}~;~m^{i}_{\alpha}= m^{i}_{M}\partial_{\alpha}X^{M}.
\end{eqnarray}

Considering the NS-NS field strength to be of the following form \cite{Bergshoeff:2015uaa},
\begin{eqnarray}
B_{MN}=\left(\omega \tau^{i}_{M} -\frac{1}{2\omega}m^{i}_{M}\right) \left(\omega \tau^{j}_{N} -\frac{1}{2\omega}m^{j}_{N}\right)\varepsilon_{ij}\label{e8}
\end{eqnarray}
it is now straightforward to obtain the \textit{finite} world-sheet action in the NC ($ \omega \rightarrow \infty $) limit,
\begin{eqnarray}
\mathcal{S}_{NG}=-\frac{\tau_{S}}{2}\int d^2 \xi ~ \sqrt{-| \mathfrak{a}|}~\mathfrak{a}^{\lambda \sigma}\mathfrak{H}_{\lambda \sigma}\label{e9}
\end{eqnarray}
where we introduce $ 2 \times 2 $ matrices, $ \mathfrak{a}_{\alpha \beta}=\tau^{i}_{\alpha}\tau^{j}_{\beta}\eta_{ij} $ together with the inverse $ \mathfrak{a}_{\alpha \beta}\mathfrak{a}^{\beta \gamma}=\delta_{\alpha}^{\gamma} $\cite{Andringa:2012uz}.

In order to proceed further, as a first step, we define matrix valued one form\footnote{We introduce matrix valued field, $ \mathfrak{j}=g^{-1}dg $ as Maurer-Cartan form \cite{Grosvenor:2017dfs} in the context of nonrelativistic strings propagating over stringy NC geometry. This could be thought of as being the generalization from centrally extended Bargmann algebra to string Galilean algebra. Here, $ g $ is an element of group $ \mathcal{G} =\exp \mathfrak{g}$ where, $ \mathfrak{g} \equiv \lbrace \mathcal{H}_{i},M_{ij},\mathcal{Z}_{i},\mathcal{Z}_{ij},G_{ip},\mathcal{P}_{q},J_{pq}\rbrace$ is a finite dimensional Lie algebra \cite{Grosvenor:2017dfs} spanned by the generators of extended string Galilean algebra \cite{Andringa:2012uz},\cite{Bergshoeff:2018yvt}. Here, $ M_{ij} $s are the generators of longitudinal Lorentz rotations and $ J_{pq} $s are the generators corresponding to transverse rotations. On the other hand, $\mathcal{Z}_{i}  $ and $ \mathcal{Z}_{ij} $ are the generators correponding non central extension of the string Galilean algebra. For details on this extended Lie algebra structure the enthusiatic reader is referred to \cite{Andringa:2012uz}.  Following \cite{Grosvenor:2017dfs}, we introduce $ \lbrace \tau^{i}_{M},m^{i}_{M},e^{q}_{M}\rbrace $ as vielbein fields associated with the coset space, $ \mathfrak{m}= \mathfrak{g}/\mathfrak{h}$ where, $ \mathfrak{h}=\lbrace M_{ij},\mathcal{Z}_{ij}, G_{ip},J_{pq}\rbrace $. Therefore, for the above coset representative the Maurer-Cartan one form becomes \cite{Grosvenor:2017dfs}, $$ \mathfrak{j}=g^{-1}dg= \mathcal{H}_{i}\tau^{i}+\mathcal{Z}_{i}m^{i}+\mathcal{P}_{q}e^{q}.$$  For a special choice of these background fields namely, $ \tau^{i}_{M}=\delta^{i}_{M},m^{i}_{M}=0,e^{q}_{M} =\delta^{q}_{M}$ the corresponding target space geometry essentially boils down to a \textit{flat} NC background \cite{Bergshoeff:2018yvt} where the question regarding integrability becomes trivial.} associated with the longitudinal translation, transverse translation as well as the non central extension of string Galilean algebra. This one form could be formally splitted into different components as,
\begin{eqnarray}
\mathfrak{j}_{\alpha}^{(\tau)}&:=&\mathfrak{j}_{\alpha}^{(\tau)i}\mathcal{H}_{i}~;~\mathfrak{j}_{\alpha}^{(\tau)i}=\tau_{M}^{i}\partial_{\alpha}X^{M}\nonumber\\
\mathfrak{j}_{\alpha}^{(e)}&:=&\mathfrak{j}_{\alpha}^{(e)q}\mathcal{P}_{q}~;~\mathfrak{j}_{\alpha}^{(e)q}=e_{M}^{q}\partial_{\alpha}X^{M}\nonumber\\
\mathfrak{j}_{\alpha}^{(m)}&:=&\mathfrak{j}_{\alpha}^{(m)i}\mathcal{Z}_{i}~;~\mathfrak{j}_{\alpha}^{(m)i}=m_{M}^{i}\partial_{\alpha}X^{M}\label{e10}
\end{eqnarray}
where, we identify $ \mathcal{H}_{i}(i=0,1) $ and $ \mathcal{P}_{q}(q=2,..,d) $ respectively as generators corresponding to longitudinal as well as transverse translations \cite{Bergshoeff:2018yvt} and $ \mathcal{Z}_{i} $ as the generator corresponding to the non central extension of string Galilei algebra \cite{Bergshoeff:2018yvt},
\begin{eqnarray}
[G_{ip},\mathcal{P}_{q}]=\delta_{pq}\mathcal{Z}_{i}
\end{eqnarray}
where, $ G_{ip} $ is the generator associated with string Galilei boosts.

From, (\ref{e10}), it immediately follows,
\begin{eqnarray}
d \mathfrak{j}+ \mathfrak{j}\wedge \mathfrak{j}=0
\end{eqnarray}
which yields,
\begin{eqnarray}
\partial_{[ M}\mathfrak{x}^{a}_{N]}=0\label{ee13}
\end{eqnarray}
where, we identify the one form in its most generic form as,
\begin{eqnarray}
\mathfrak{j}^{(\mathfrak{x}_{a})}_{\alpha}=\mathfrak{x}_{M}^{a}\partial_{\alpha}X^{M}T_{a}=\mathfrak{j}_{\alpha}^{(\mathfrak{x}_{a})a}T_{a}.\label{e14}
\end{eqnarray}
Here, $ T_a (=\mathcal{H}_{i},\mathcal{Z}_{i},\mathcal{P}_{q}) $ could be identified with one of the \textit{translation} generators (\ref{e10}) that satisfies \cite{Andringa:2012uz},
\begin{eqnarray}
[ T_a , T_b]=0.
\end{eqnarray}

With the above terminology in place, it is now quite straightforward to re-express the NG action (\ref{e9}) in terms of (\ref{e10}) defined above,
\begin{eqnarray}
\mathcal{S}_{NG}=-\tau_{S}\int d^2 \xi ~ \sqrt{-| \mathfrak{a}|}~\mathfrak{a}^{\alpha \beta}~\mathfrak{j}_{\alpha}^{(\mathfrak{x}_{a})a}~\vartheta_{ab}~\mathfrak{j}_{\beta}^{(\mathfrak{z}_{b})b}
\end{eqnarray}
where, we introduce the following entities,
\begin{eqnarray}
\mathfrak{x}_{a}&=& e \delta_{pa}+\tau \delta_{ia}~;~\mathfrak{z}_{b}= e \delta_{qb}+m \delta_{jb}\nonumber\\
\vartheta_{pq}&=&\delta_{pq}~;~\vartheta_{ij}=\eta_{ij}~;~a=i,p~;~b=j,q\label{e13}
\end{eqnarray}
where, $ i,j $ are the longitudinal indices and $ p,q $ are the transverse indices defined in (\ref{e3}).

Next, we note down the equation of motion corresponding to $ \mathfrak{j}^{(\mathfrak{x}_{a})a} $,
\begin{eqnarray}
\partial_{\alpha}(\sqrt{-| \mathfrak{a}|}~\mathfrak{a}^{\alpha \beta}~\Im_{\beta a}) =0\label{e16}
\end{eqnarray}
where, we introduce new variables,
\begin{eqnarray}
\Im_{\beta a}&=&\mathfrak{j}_{\beta}^{(\mathfrak{z}_{b})b}\vartheta_{ab} + \mathfrak{j}_{\beta}^{(e)p} \delta_{p a} +\varpi (\mathfrak{j})~\mathfrak{j}_{\beta}^{(\tau)i}~\eta_{ia}-\mathfrak{a}^{\lambda \gamma}\left(\Xi^{i}_{\beta \lambda \gamma}+\Xi^{i}_{\lambda \beta\gamma } \right)\eta_{ia} \nonumber\\
\varpi (\mathfrak{j}) &=& \mathfrak{a}^{\alpha \beta}~\mathfrak{j}_{\alpha}^{(\mathfrak{x}_{a})a}~\mathfrak{j}_{\beta}^{(\mathfrak{z}_{b})b}\vartheta_{ab}~;~
\Xi^{i}_{\beta \lambda \gamma}=\mathfrak{j}_{\beta}^{(\mathfrak{x}_{\tilde{a}})\tilde{a}}~\mathfrak{j}_{\lambda}^{(\mathfrak{z}_{\tilde{b}})\tilde{b}}~\mathfrak{j}_{\gamma}^{(\tau)i}~\vartheta_{\tilde{a}\tilde{b}}~;~\tilde{a}=i,p~;~\tilde{b}=j,q.\label{ne17}
\end{eqnarray}

Finally, we note down the equation of motion corresponding to $ \mathfrak{j}^{(\mathfrak{z}_{b})b} $,
\begin{eqnarray}
\partial_{\alpha}\left(\sqrt{-|\mathfrak{a}|}\mathfrak{a}^{\alpha \beta}(\mathfrak{j}_{\beta}^{(\mathfrak{x}_{a})a}\vartheta_{ab}+\mathfrak{j}_{\beta}^{(e)q}\delta_{bq}) \right)=0.\label{ne18} 
\end{eqnarray} 

Based on (\ref{ee13}), (\ref{e16}) and (\ref{ne18}), we propose the following \textit{flat} (Lax) connection associated with string NC geometry,
\begin{eqnarray}
\mathfrak{L}_{\alpha}=\mathfrak{L}^{a}_{\alpha}~T_a
\end{eqnarray}
where, we express the individual coefficients as,
\begin{eqnarray}
\mathfrak{L}^{a}_{\alpha}&=&\frac{1}{1-\Lambda^{2}}\left(\mathfrak{j}_{\alpha}^{(\mathfrak{x}_{a})a}-\Lambda \varepsilon_{\alpha \beta}\sqrt{-| \mathfrak{a}|}\mathfrak{a}^{ \beta \gamma} \Re_{ \gamma}^{a}\right)\nonumber\\
\Re_{ \beta a} & =& \sum_{c=0}^{d}(\mathfrak{j}^{(\mathfrak{x}_{c})c}_{\beta}+\mathfrak{j}^{(\mathfrak{z}_{c})c}_{\beta})\vartheta_{ac}+2\sum_{p=2}^{d}\mathfrak{j}_{\beta}^{(e)p}\delta_{pa}+\sum_{i=0,1}\left(\varpi (\mathfrak{j})\mathfrak{j}_{\beta}^{(\tau)i}-\mathfrak{a}^{\lambda \gamma}\left(\Xi^{i}_{\beta \lambda \gamma}+\Xi^{i}_{\lambda \beta\gamma } \right)\right) \eta_{ia}\nonumber\\   \label{e17}
\end{eqnarray}
where $ \Lambda $ is the so called spectral parameter together with, $ \varepsilon_{\sigma^{0} \sigma^{1}}=-\varepsilon_{\sigma^{1} \sigma^{0}}=1 $.

Using (\ref{ee13}), (\ref{e16}) and (\ref{ne18}) it is now trivial to show,
\begin{eqnarray}
\varepsilon^{\alpha \beta}\partial_{\alpha}\mathfrak{L}_{\beta}=\partial_{\alpha}\mathfrak{L}_{\beta}-\partial_{\beta}\mathfrak{L}_{\alpha}=0.
\end{eqnarray}

On the other hand, a careful computation reveals,
\begin{eqnarray}
\left[\mathfrak{L}_{\alpha},\mathfrak{L}_{\beta} \right]=\mathfrak{L}^{a}_{\alpha}\mathfrak{L}^{b}_{\beta}\left[T_a, T_b \right]=0~;~a,b=i,q
\end{eqnarray}
where, by virtue of \textit{extended} string Galilei algebra \cite{Brugues:2004an}-\cite{Andringa:2012uz}, we identify each of the above commutators equal to zero. This finally allows us to write down the zero curvature condition for Lax pairs in two dimensions \cite{Arutyunov:2009ga},
\begin{eqnarray}
\partial_{\alpha}\mathfrak{L}_{\beta}-\partial_{\beta}\mathfrak{L}_{\alpha}-\left[\mathfrak{L}_{\alpha},\mathfrak{L}_{\beta} \right]=0. 
\end{eqnarray}
\section{Conserved charges}
The flat connection (\ref{e17}) provides us a canonical way of estimating conserved charges associated with the 2D sigma model. The first step is to write down the corresponding \textit{monodromy} matrix \cite{Arutyunov:2009ga},
\begin{eqnarray}
\mathfrak{T}(\Lambda)=P~\exp \int_{0}^{2\pi}d \sigma^{1} \mathfrak{L}_{\sigma^{1}}(\Lambda):=\exp \mathcal{A}(\Lambda)\label{e21}
\end{eqnarray}
where, the exponential is subjected to a path ordered form. 

Using (\ref{e17}), one could further simplify (\ref{e21}) as\footnote{See Appendix for the details.},
\begin{eqnarray}
\mathcal{A}(\Lambda)= \int_{0}^{2\pi}d \sigma^{1} ~ \mathfrak{L}^{a}_{\sigma^{1}}(\Lambda)T_{a}= \int_{0}^{2\pi}d \sigma^{1} ~\Gamma^{b}(\alpha , \Lambda)~\vartheta_{ba}~T_a \label{e22}
\end{eqnarray}
where, we assume that all the \textit{dynamical} variables associated with the world-sheet theory are \textit{periodic} functions of $ \sigma $.

Using (\ref{e22}), it is now straightforward to construct the corresponding \textit{transfer} matrix, 
\begin{eqnarray}
\mathfrak{t}(\Lambda)= tr~ \mathfrak{T}(\Lambda)=tr~ e^{\int_{0}^{2 \pi}d \sigma^{1} \Gamma^{b}(\alpha , \Lambda)\vartheta_{ba}T_{a}}.\label{e29}
\end{eqnarray}

A straightforward computation reveals \cite{Arutyunov:2009ga},
\begin{eqnarray}
\partial_{0}\mathfrak{t}=[\mathfrak{L}_{\sigma^{0}}(\sigma^{0},\Lambda),\mathfrak{T}(\Lambda)]=0
\end{eqnarray}
which eventually leads to an infinite tower of \textit{conserved} charges $ \mathcal{Q}_{n} $  ($ \forall n \geq 0 $),
\begin{eqnarray}
\mathfrak{t}(\Lambda)=\sum_{n=0}^{\infty}\mathcal{Q}_{n}~\Lambda^{n}
\end{eqnarray} 
thereby making the theory integrable. By Taylor expanding the transfer matrix (\ref{e29}) near the origin we finally read off the family of non-local conserved charges,
\begin{eqnarray}
\mathcal{Q}_{n} =\frac{1}{n !}\frac{\partial^{n}}{\partial \Lambda^{n}}~tr \sum_{m,k=0}^{\infty}\frac{1}{k !}\left(\int_{0}^{2 \pi}d\sigma^{1}\frac{\Lambda^{m}}{m!}\frac{\partial^{m}}{\partial \Lambda^{m}}\Gamma^{b}(\alpha , \Lambda)\vartheta_{ba}T_{a} \right)^{k}\Big|_{\Lambda =0}
\end{eqnarray}
associated with the 2D non-linear sigma model corresponding to stringy NC background. 
\section{Summary and final remarks}
The present paper was an attempt to explore classical integrability for non-relativistic strings propagating over stringy Newton-Cartan (NC) background. We address this issue considering the bosonic sector of the full superstring spectra. The analysis of this paper suggests that the non-linear sigma model under consideration indeed allows an underlying integrable structure that further opens up a number of possible future directions. The first and foremost question that one should clarify is whether the integrable structure could be extended for the full superstring spectrum. The other question that remains to be addressed is what are the natural consequences of the above integrable structure in the context of AdS/CFT duality. In order to address the second question, one must have a detailed understanding of the stringy dynamics in the presence of a negative cosmological constant \cite{Andringa:2012uz}. We leave these issues for the purpose of future investigations.\\ \\ 
{\bf {Acknowledgements :}}
 The author is indebted to the authorities of IIT Roorkee for their unconditional support towards researches in
basic sciences. \\\\ 
{ \large{ {\bf {Appendix: Expression for $ \Gamma^{b}(\alpha , \Lambda) $}}}}\\
Here, in the Appendix, we provide the detail expression for the function $\Gamma^{b}(\alpha , \Lambda)  $ appearing in the expression for the monodromy matrix (\ref{e21}),
 \begin{eqnarray*}
\Gamma^{b}(\alpha , \Lambda)=~~~~~~~~~~~~~~~~~~~~~~~~~~~~~~~~~~~~~~~~~~~~~~~~~~~~~~~~~~~~~~~~~~~~~~~~~~
~~~~~~~~~~~~~~~~~~~~~~~~~~~~~~~~~~~~~~~~~~~~~~~\nonumber\\
\frac{\Lambda}{1-\Lambda^{2}}~\sum_{\mathfrak{m}=\mathfrak{x},\mathfrak{z}}\left[\sqrt{-| \mathfrak{a}|}\mathfrak{a}^{00}(\mathfrak{j}_{0}^{(\mathfrak{m})b}+2\mathfrak{j}_{0}^{(e)p}\delta_{p}^{b})-\partial_{1}\left( \sqrt{-| \mathfrak{a}|}\mathfrak{a}^{01}\mathfrak{m}_{N}^{b}\right)X^N -\partial_{1}\left( \sqrt{-| \mathfrak{a}|}\mathfrak{a}^{01}e^{p}_{N}\delta_{p}^{b}\right)X^N\right]~~~~~~~~~~~~~\nonumber\\
-\frac{1}{1-\Lambda^{2}}\partial_{N}\mathfrak{x}^{b}_{M}\partial_1 X^N X^M+\frac{\Lambda}{1-\Lambda^{2}}~\left[  \sqrt{-| \mathfrak{a}|}\mathfrak{a}^{00}\left( \varpi(\mathfrak{j})\mathfrak{j}_{0}^{(\tau)i}-\left( \mathfrak{a}^{\alpha \beta}\mathfrak{j}_{0}^{(\mathfrak{x}_{\tilde{a}})\tilde{a}}\mathfrak{j}_{\alpha}^{(\mathfrak{z}_{\tilde{b}})\tilde{b}}\mathfrak{j}_{\beta}^{(\tau)i}\vartheta_{\tilde{a}\tilde{b}}+(\mathfrak{x}\leftrightarrow \mathfrak{z})\right) \right) \right]\delta^{b}_{i}~~~~~~~~~~~\nonumber\\
+\frac{\Lambda}{1-\Lambda^{2}}~\left[ \left(  \partial_{1}\left( \sqrt{-| \mathfrak{a}|}\mathfrak{a}^{01}\mathfrak{a}^{\alpha \beta}\mathfrak{j}_{\alpha}^{(\mathfrak{z}_{\tilde{b}})\tilde{b}}\mathfrak{j}_{\beta}^{(\tau)i}\mathfrak{x}_{M}^{\tilde{a}}\vartheta_{\tilde{a}\tilde{b}}\right)X^M+ (\mathfrak{x}\leftrightarrow \mathfrak{z})\right) -\partial_{1}\left( \sqrt{-| \mathfrak{a}|}\mathfrak{a}^{01}\varpi (\mathfrak{j})\tau_{M}^{i}\right) X^M \right]\delta^{b}_{i}~~~~~~~~~~
\end{eqnarray*}
where we use short-hand notation for the derivative, $ \frac{\partial}{\partial\sigma^{\alpha}}\equiv \partial_{\alpha} $.

\end{document}